\documentclass[conference]{IEEEtran}

\usepackage[utf8]{inputenc}        
\usepackage[scr=boondoxo,frak=euler,bb=ams]{mathalfa}
\usepackage{amsmath,amsthm,amsfonts,amssymb}     
\usepackage{graphicx,graphics}
\usepackage{float}
\usepackage{tikz}
\usetikzlibrary{shapes, snakes, patterns}
\usepackage{xcolor}        

\usepackage[export]{adjustbox}
\usepackage[justification=centering]{caption}
\usepackage{subcaption}
\usepackage{placeins}
\DeclareCaptionLabelSeparator{periodspace}{.\quad}
\usepackage{epstopdf}
\usepackage{enumerate,enumitem} 
\usepackage{cite}
\usepackage{suffix}
\usepackage{mathtools}
\usepackage{tabularx}
\usepackage{booktabs}        
\usepackage{boldline,multirow,diagbox,hhline} 
\usepackage{slashbox}
\usepackage{url} 
\usepackage{algorithm,algpseudocode}
\usepackage{relsize}
\usepackage{cuted}
\usepackage{titlesec}
\usepackage{chngcntr}
\usepackage{multicol}
\usepackage{textcomp}
\def\BibTeX{{\rm B\kern-.05em{\sc i\kern-.025em b}\kern-.08em
		T\kern-.1667em\lower.7ex\hbox{E}\kern-.125emX}}

\theoremstyle{definition}

\newtheorem{theorem}{Theorem}[]

\newcommand{\PreserveBackslash}[1]{\let\temp=\\#1\let\\=\temp}
\newcolumntype{C}[1]{>{\PreserveBackslash\centering}p{#1}} 
\newcolumntype{R}[1]{>{\PreserveBackslash\raggedleft}p{#1}} 
\newcolumntype{L}[1]{>{\PreserveBackslash\raggedright}p{#1}} 

\newcolumntype{Y}{>{\centering\arraybackslash}b{0.7cm}}
\newcolumntype{M}{>{\centering\arraybackslash}b{2.1cm}}
\newcolumntype{Z}{>{\centering\arraybackslash}b{1.75cm}}
\newcolumntype{G}{>{\centering\arraybackslash}m{1.5cm}}
\newcolumntype{Q}{>{\centering\arraybackslash}b{3.75cm}}
\newcolumntype{A}{>{\centering\arraybackslash}m{1.5cm}}
\newcolumntype{B}{>{\centering\arraybackslash}m{0.6667cm}}
\newcolumntype{T}{>{\centering\arraybackslash}b{6.5ex}}
\newcolumntype{S}{>{\centering\arraybackslash}b{13.25ex}}
\newcolumntype{?}{!{\vrule width 1pt}}
\newcolumntype{+}{!{\vrule width 2pt}}

\DeclareFontFamily{U}{mathx}{}
\DeclareFontShape{U}{mathx}{m}{n}{<-> mathx10}{}
\DeclareSymbolFont{mathx}{U}{mathx}{m}{n}
\DeclareMathAccent{\widecheck}{0}{mathx}{"71}

\definecolor{bluee}{RGB}{0, 82, 200}
\definecolor{redd}{RGB}{245, 30, 30}
\definecolor{greenn}{RGB}{0, 150, 62}

\makeatletter
\def\bigc#1{{\hbox{$\left#1\vbox to0.5\p@{}\right.\n@space$}}}
\makeatother

\newcommand{\mbf}[1]{\boldsymbol{\mathrm{#1}}}

\newcommand{\norm}[1]{\left\lVert#1\right\rVert}

\newcommand{\hyph}{$\psp$-$\psp$}

\newcommand{\commhide}[1]{}

\newcommand{\psp}{\hspace{0.1em}}
\newcommand{\pspp}{\hspace{0.05em}}
\newcommand{\psppp}{\hspace{0.065em}}
\newcommand{\nsp}{\hspace{-0.1em}}
\newcommand{\nspp}{\hspace{-0.05em}}

\makeatletter
\newcommand{\vast}{\bBigg@{4}} 
\newcommand{\Vast}{\bBigg@{5}} 
\newcommand{\csize}[1]{\bBigg@{#1}} 
\makeatother

\makeatletter
\newcommand{\ostar}{\mathbin{\mathpalette\make@circled\star}}
\newcommand{\make@circled}[2]{\ooalign{$\m@th#1\smallbigcirc{#1}$\cr\hidewidth$\m@th#1#2$\hidewidth\cr}}
\newcommand{\smallbigcirc}[1]{ \vcenter{\hbox{\scalebox{0.77778}{$\m@th#1\bigcirc$}}}}
\makeatother

\newcommand{\range}{\mathscr{R}}
\newcommand{\kernel}{\mathscr{K}}

\newcommand{\diag}{\operatorname{diag}}

\newcommand{\herm}[1]{{#1}^{\dagger}_{}}
\newcommand{\hermm}[1]{{#1}^{\dagger}}

\newcommand{\summ}{\textstyle\sum\limits}

\title{Low-Complexity Linear 
	Decoupling of Users for Uplink 
	Massive MU-MIMO 
	Detection}

\author{
	\IEEEauthorblockN{%
		S. Sowmya\textsuperscript{*}, %
		$\ $%
		Gokularam Muthukrishnan\textsuperscript{*}, %
		$\ $%
		K. Giridhar
		}
	\IEEEauthorblockA{\textit{TelWiSe Group, Department of Electrical Engineering,} \\
		\textit{Indian Institute of Technology Madras, 
		Chennai - 600036, India.
		}\\
		e-mail: \{sowmya,$\,$gokularam,$\,$k.giridhar\}@telwise-research.com}
}

\begin{document}

\maketitle

\begingroup\renewcommand\thefootnote{*}
\footnotetext{Equal contribution.}
\endgroup

\begin{abstract}

 Massive MIMO (mMIMO) 
  enables users with different requirements to get connected to the same base station (BS) on the same set of resources. In the uplink of
  Multi-user massive MIMO (MU-mMIMO), while such heterogeneous users are served, 
  decoupling facilitates the use of user-specific detection schemes. In this paper, we propose a low-complexity linear decoupling scheme called Sequential Decoupler (SD), which aids in the parallel detection of each user's data stream. The proposed algorithm shows significant complexity reduction. 
  Simulations reveal that the complexity of the proposed scheme is only {$0.15\%$} of the conventional Singular Value Decomposition (SVD) based decoupling and is about $47\%$ of the pseudo-inverse based decoupling schemes when 80 users with two antennas each are served by the BS. 
  Also, the proposed scheme is scalable when new users are added to the system 
  %
  and requires 
  fewer operations than computing the decoupler all over again.
  %
  %
  Further numerical analyses indicate that the proposed scheme achieves significant complexity reduction without any degradation in performance and is a promising low-complex alternative to the existing decoupling schemes.
 
\end{abstract}

\section{Introduction}
Massive Multiple Input Multiple Output (mMIMO) communication offers several advantages, such as increased spectral efficiency and reliability, and is a suitable candidate for the next generation wireless networks 
\cite{larsson2014massive,adhikari20226g}. 
It provides improved utilization of radio resources by receiving and transmitting  signals at the same time and frequency resources by exploiting the spatial degrees of freedom. In uplink multi-user massive  MIMO (MU-mMIMO), the base station (BS) receives signals from several users simultaneously, and all users' signals need to be decoded. 
Typically, in MU-mMIMO scenarios, BS will have more antennas than user terminals. 
In order to make the user terminal design simple, the issue of handling the users' interference is vested with the BS. Thus, the requirement of computationally efficient interference handling schemes with multiple users having different requirements is of key importance.

\par 
The optimal Maximum Likelihood (ML) detection will give the best performance but is not an option even for small-scale MIMO systems as the computational complexity 
scales exponentially with 
the number of user streams to be decoded. 
Linear inverse channel detection schemes such as Zero-Forcing (ZF) detectors and Linear Minimum Mean Squared Error (LMMSE) detectors offer low complexity detection but at the cost of performance degradation \cite{paulraj2003introduction}. 
Non-linear sub-optimal schemes such as Successive Interference Cancellation (SIC) \cite{wubben2001efficient} offer improved performance over linear schemes. SIC sequentially decodes symbols and excises their impact on the received signal for successive detections. However, with improper decoding order, 
the performance of SIC can degrade 
due to error propagation. Several schemes 
\cite{wolniansky1998v,wubben2001efficient,stankovic2005improved,li2011multiple,de2013adaptive} have been devised to identify an appropriate ordering for SIC, but they impose 
very high 
computational 
costs, particularly for 
large-scale massive MIMO systems. 
Also, the aforementioned schemes do not provide the 
elegance and ease of using independent detectors for each user.

\par
Decoupled Detection (DD) \cite{arevalo2015uplink,arevalo2017decoupled,zu2019uplink} 
enables parallel detection by segregating users (or groups of users) and using the existing schemes to detect the segregated streams with lesser complexity,  
and the
segregation is achieved through linear 
decouplers. The design of such decouplers is
inspired by the existing 
precoding schemes for the downlink \cite{arevalo2015uplink}. Thus, DD schemes facilitate the  \textit{hybridization} of linear and non-linear detection methods, 
enabling a distinct trade-off between 
the performance and computational cost. 
They also facilitate the use of user-specific detectors independent of other users since every user might have their own requirements.
\subsection{Prior Works and Motivation}
%
%
%
%

\par
Authors of \cite{arevalo2015uplink} proposed a decoupler for each user that projects the received signal onto the left nullspace of 
its complementary channel, which is computed using the Singular Value Decomposition (SVD) before detecting 
its symbols. 
It has to be noted that the decoupler has to be computed for every user, and thus, the computational complexity of this scheme is high. In \cite{arevalo2017decoupled} and \cite{zu2019uplink}, the authors proposed a scheme that 
computes the decoupler using the pseudo-inverse of the channel matrix. 
These DD schemes have been used extensively for parallel detection on the uplink \cite{tseng2023hybrid}. 
The main disadvantage of the aforementioned schemes is that once there is a 
new user in the system, the decoupler of every user has to be recomputed. 
Motivated by this, we have designed  a low-complexity decoupling scheme that can adapt to the changes in the system. 
\subsection{Contributions}
\begin{enumerate}[label=(\roman*)]
	\item We propose a novel, low-complex linear decoupler called Sequential Decoupler (SD), which 
	aids in processing every user's streams individually.
	
	
	\item The proposed decoupler is scalable and helps in the inclusion of new users into the system 
	without having to recompute all over again. 
	
	\item 
	
	Extensive complexity and Bit Error Rate (BER) 
	analyses corroborate 	that the proposed scheme achieves remarkable complexity reduction without any performance degradation.
\end{enumerate}
To the authors' best knowledge, the proposed sequential decoupler is not available or published in open literature.\\

\noindent\textbf{Basic Notations:} 
Boldface capital letters and boldface small letters denote matrices and vectors, respectively. The $i$-th entry of vector $\mbf{a}$ and $(i,j)$-th entry of matrix $\mbf{A}$ are respectively denoted as $(\mbf{a})_i^{}$ and $(\mbf{A})_{i,\pspp j}^{}$. The unary operator $\lceil\cdot\rceil$ denotes the \textit{ceil} function, i.e., it provides the largest integer closest to its argument. The set of all complex numbers is denoted as $\mathbb{C}$. We use  $\mathbb{N}_{{J}}^{}$ to denote the set of first $J$ positive integers, 
i.e.,  $\mathbb{N}_{{J}}^{}=\{1,\, 2, \, \ldots,\, {J}\}$. For the ordered, countable set $\mathcal{S}\pspp$, $|\mathcal{S}|$ indicates its cardinality; $\mathcal{S}[\pspp\Omega\pspp]$ denotes the subset of $\mathcal{S}$ formed from the entries that are indexed by $\Omega\pspp$, where $\Omega\subseteq\mathbb{N}_{|\mathcal{S}|}^{}$. Let $\mathcal{CN} (\mu, \sigma^2 )$ indicate circularly symmetric complex Gaussian distribution of mean $\mu$ and variance $\sigma^2$.  ${\mbf{I}_{M}^{}}$ is the identity matrix of dimension $M$. $\range(\mbf{A})$ and $\kernel(\mbf{A})$ respectively denote the range space and the 
kernel of the matrix $\mbf{A}$; $\mbf{A}^\top_{}$, 
$\herm{\mbf{A}}$ and $\norm{\mbf{A}}_{\texttt{F}}^{}$ respectively denote 
the transpose, adjoint, and Frobenius norm of the matrix $\mbf{A}\pspp$. 


\section{Background}
In this section, we will introduce 
the system model and describe the decoupled detection of user streams. 

\subsection{System Model}
We consider the uplink MU-mMIMO systems with $K$ users, 
and the $i$-th user has $M_i^{}$ transmit antennas, $i \in \mathbb{N}_K^{}$. The 
BS, which is equipped with $N_R^{}$ receiver antennas, receives data from all the users. With perfect time and frequency synchronization, the received signal at the BS is given by
\begin{equation}\label{eq:rx_sig}
	\mbf{y}=\summ_{i=1}^{K}\mbf{H}_i^{} \psp \mbf{x}_i^{} +\mbf{n}
	\psp,
\end{equation}
where $\mbf{y}\in \mathbb{C}^{N_R^{}\times 1}_{}$, $\mbf{H}_i^{}\in \mathbb{C}^{N_R^{}\times M_i^{}}_{}$ is the $i$-th user's channel, $\mbf{x}_i^{} \in \mathbb{C}^{M_i^{}\times 1}_{}$ is the $i$-th user's signal and $\mbf{n} \in \mathbb{C}^{N_R^{} \times 1}\sim \mathcal{CN}({0},  \sigma_n^2 \psp{\mbf{I}}_{{M}_i^{}}^{} \nspp)$ is the 
receiver noise at the BS.
In order to make the mobile terminals'  design simple, user terminals do not transmit signals with any pre-processing such as precoding. The BS has to 
eliminate multi-user interference in the received vector in order to process each user's signal. 

In order to facilitate inter-user interference management, 
it is assumed that the BS has good estimates of users' channels, which are obtained 
prior to the detection. 
%
%
The received signal in \eqref{eq:rx_sig} can be re-expressed as $\mbf{y}=\mbf{H}\psp\mbf{x}+\mbf{n}\pspp$, where $	\mbf{H}=\left[\psp \mbf{H}_1^{}\ \mbf{H}_2^{}\ \cdots \ \mbf{H}_K^{} \psp \right]
\in \mathbb{C}^{N_R^{} \times M}$
is the 
overall channel matrix comprising of all the users' channels, 
$M=\sum_{j=1}^{K}M_j^{}\pspp$, and $\mbf{x}=\left[\psp\mbf{x}_1^\top\ \mbf{x}_2^{\top}\ \cdots \ \mbf{x}_K^\top\psp\right]^{\top}_{}$ is the collective transmit signal to be decoded at the BS.
The complementary channel of the $i$-th user can be given by
\begin{equation}
	\overline{\mbf{H}}_i^{}=\left[\pspp \mbf{H}_1^{}\ \cdots\ \mbf{H}_{i-1}^{} \ \mbf{H}_{i+1}^{} \cdots \ \mbf{H}_{K}^{}\right]\nsp
	\in  \mathbb{C}^{N_R^{} \times \overline{M}_i^{}}_{}\pspp,
\end{equation}
where 
$\overline{M}_i^{}= 
M-M_i^{}\pspp$. 

\subsection{Decoupled Detection}
The BS has to decouple every user's signal in order to demodulate the corresponding user's streams effectively. 
Let 
\begin{equation}
	\mbf{W}=\Big[\pspp\mbf{W}_1^\top\ \mbf{W}_2^{\top}\ \cdots \ \mbf{W}_K^\top\Big]^{\top}_{}
	\in \mathbb{C}^{M\times N_R^{}}_{}
\end{equation}
denote the linear decoupler, where $\mbf{W}_i^{}\in\mathbb{C}^{M_i^{}\times N_R^{}}_{}\pspp$ is the decoupler corresponding to the $i$-th user, $i\in\mathbb{N}_{K}^{}\pspp$.
The linear decoupling is performed by satisfying (or nearly satisfying) the condition  ${\mbf{W}_i^{}}\overline{\mbf{H}}_i^{}=\mbf{0}$ \cite{arevalo2015uplink}; i.e., the rows of $\mbf{W}_i^{}$ 
have to be in the left nullspace of the user's complementary channel, $\overline{\mbf{H}}_i^{}\pspp$. 
By applying the decoupler to the received signal $\mbf{y}\pspp$, i.e., $\widetilde{\mbf{y}}=\mbf{W}\mbf{y}\pspp$, 
the uplink channel is transformed into  a set of parallel single-user MIMO channels, and user-specific detection schemes can be applied on the slices of $\widetilde{\mbf{y}}$ corresponding to each user.

With the requirement that $\overline{M}_i^{}<N_R^{}\pspp$, the left nullspace 
can be computed using SVD of the user's complementary channel, $\overline{\mbf{H}}_i^{}$,
\begin{equation}
	\overline{\mbf{H}}_i^{}=\big[\psp \overline{\mbf{U}}_i^{{}_{(1)}} \ \overline{\mbf{U}}_i^{{}_{(0)}}\psp \big] \ \overline{\mbf{\Sigma}}_i^{} \ \hermm{\overline{\mbf{V}}}_i
	\pspp;
\end{equation}
the columns of $\mbf{\overline{U}}_i^{{}_{(0)}}\in \mathbb{C}^{N_R^{} \times (N_R^{}-\overline{M}_i^{})}_{}$
will constitute a basis for the left nullspace of $\mbf{\overline{H}}_i^{}$. Thus, $\mbf{W}_i^{}={\big(\mbf{\overline{U}}_i^{{}_{(0)}}\big)^{\dagger}}\psp$ is used as a decoupler in \cite{arevalo2015uplink}. However, different decouplers can be obtained by choosing different basis spanning (or approximately spanning) the left nullspaces of the complementary channels.

Since ${\mbf{W}_i^{}}\overline{\mbf{H}}_i^{}=\mbf{0} \iff \mbf{W}_i^{}\mbf{H}_k^{}=\mbf{0} \,\ \forall\, k \psp \in \psp \mathbb{N}_{K}^{}\nspp \backslash \{i\}\pspp$, the decouplers need to satisfy the condition
\begin{equation}\label{eq:LNS_intersection}
	\range(\hermm{\mbf{W}}_i)
	=
	\bigcap\limits_{k \psp \in \psp \mathbb{N}_{K}^{}\nspp \backslash \{i\} }^{}{\kernel(\hermm{\mbf{H}}_k)}
	\pspp,
	\quad
	i\in\mathbb{N}_K^{}
	\pspp,
\end{equation}
i.e., each user's decoupler lies in the intersection of left nullspaces of 
all other users'
channel matrices. 
Thus, computing the decouplers in the aforementioned way involves a lot of redundancy. 
A low-complexity decoupler design is proposed in the following section by harnessing this redundancy.

\subsection{Detection of User Streams}{\label{Detection_schems}
As mentioned before, 
each user's data can be individually detected after decoupling their streams. 
%
Consider the decoupler $\mbf{W}_i^{}$ of user $i\in\mathbb{N}_K^{}$; once the decoupler is applied to the received signal $\mbf{y}$, the 
resulting signal $\widetilde{\mbf{y}}_i^{}=\mbf{W}_i^{}\psp\mbf{y}$ corresponds to the received signal arising from the user's data stream $\mbf{x}_i^{}$ passing through the equivalent channel $\widetilde{\mbf{H}}_{i}^{}=\mbf{W}_i^{}\mbf{H}_i^{} 
\pspp$, corrupted by the additive noise $\widetilde{\mbf{n}}_i^{}=\mbf{W}_i^{}\psp\mbf{n} \pspp$. 
Any detection algorithm to recover $\mbf{x}_i^{}$ can be applied on $\widetilde{\mbf{y}}_i^{}\pspp$. 
Thus, decoupling reduces the 
system dimension, facilitating the use of complex non-linear detection schemes for each user.  Also, the user streams can be detected parallelly after decoupling. We consider the following detection schemes in our work.  
\subsubsection{LMMSE detector \cite{paulraj2003introduction}:}
In the LMMSE detector, the received stream is passed through a linear filter to eliminate the channel effects before the demodulation.  From the equivalent received signal $\mbf{y}_i^{}$, the $i$-th user stream is detected by demodulating each coordinate of $\widehat{\mbf{x}}_i^{}=\mbf{G}_i^{}\psp\mbf{y}_i^{}$, where $\mbf{G}_{i}^{}=\Big(\hermm{\widetilde{\mbf{H}}_{i}}\widetilde{\mbf{H}}_{i}^{}+\sigma_n^2\mbf{I}\pspp\Big)^{-1}\hermm{\widetilde{\mbf{H}}_{i}}\in \mathbb{C}^{M_i^{}\times M_i^{}}_{}.$
\subsubsection{Successive Interference Cancellation (SIC) \cite{wubben2001efficient}:}
The non-linear SIC detectors recursively eliminate inter-symbol interference within the user streams by demodulating one symbol at a time. 
In this work, we consider QR decomposition-based SIC. Consider the QR decomposition of the $i$-th user's effective channel $\widetilde{\mbf{H}}_{i}^{}=\mbf{Q}_i^{}\mbf{R}_i^{}$ and let $\mbf{v}_i^{}=\hermm{\mbf{Q}}_i\widetilde{\mbf{y}}_i^{}$. SIC detection is performed as
\begin{equation*}
	[\widehat{\mbf{x}}_i^{}]_j^{} = \mathcal{Q}\!\left(\nsp\frac{(\mbf{v}_i^{})_j^{}-\sum_{p=j+1}^{M_i^{}}(\mbf{R}_i^{})_{j,\pspp p}^{}	(\widehat{\mbf{x}}_i^{})_p^{}}{(\mbf{R}_i^{})_{j,\pspp j}^{}}\nsp\right)\!
	\psp,
	\end{equation*}
	where $j$ is decreased from $M_i^{}$ to $1$ and $\mathcal{Q}(\cdot)$ indicates the slicing function that maps its input to the nearest alphabet. Typically, SIC achieves better performance than LMMSE owing to the fact that the SNR gets maximized every time a stream is eliminated recursively \cite{wubben2001efficient}.

\section{Proposed Sequential Decoupler (SD)}
We now introduce our proposed approach for computing the decoupler.
We first describe a recursive procedure to estimate the left nullspace\footnote{In the remainder of the article, with a slight abuse of terminology, we 
call the matrix $\mbf{W}$ comprising of rows which are the basis of the left nullspace of $\mbf{A}$ simply as the left nullspace of $\mbf{A}\pspp$.} of a matrix with column partitions, which
will help us find a low-complex decoupler by exploiting the redundancy in \eqref{eq:LNS_intersection}.

\subsection{Recursive Estimation of Left Nullspace}
The following result, 
inspired from \cite[Theorem 6.4.1.]{golub2013matrix},  %
presents a way to compute the common left nullspace of two matrices.
\begin{theorem}
	\label{thm:IntersectionofNullSpaces}		
	Consider the matrices $\mbf{A}\in \mathbb{C}^{n\times m}_{}$ and $\mbf{B}\in \mathbb{C}^{n\times p}_{}$. Let the rows of $\mbf{F} \in \mathbb{C}^{t \times n}_{}$ constitute the basis for $\kernel(\herm{\mbf{A}})$. Also, let the rows of $\mbf{G} \in \mathbb{C}^{q \times t}_{}$ constitute the basis for $\kernel\big(\herm{(\mbf{F}\pspp\mbf{B})}\big)$. Then, rows of ${\mbf{G}\pspp \mbf{F}}$ form a basis for ${\kernel(\herm{\mbf{A}})}\psp \cap \psp {\kernel(\herm{\mbf{B}})}$.
	
	\begin{proof}
		We have $\mbf{F}\pspp\mbf{A}=\mbf{0}
		\implies 
		\mbf{G}\pspp\mbf{F}\pspp\mbf{A}=\mbf{0}\pspp$. 
		Also, we have $\mbf{G}\pspp\mbf{F}\pspp\mbf{B}=\mbf{0}\pspp$. Therefore, $\range\big(\herm{(\mbf{G}\pspp\mbf{F})}\big)
		\subseteq
		 {\kernel(\herm{\mbf{A}})}\psp \cap \psp {\kernel(\herm{\mbf{B}})}\pspp$.
		Now, let us consider an arbitrary vector $\mbf{x} \in  {\kernel(\herm{\mbf{A}})}\psp \cap \psp {\kernel(\herm{\mbf{B}})}\pspp$. 
		Since $\mbf{x} \in {\kernel(\herm{\mbf{A}})}\pspp$, 
		$\exists \ { \mbf{y} \in}\ \mathbb{C}^{t\times 1}_{}$ such that ${\mbf{x}}={\herm{\mbf{F}}\mbf{y}}\pspp$.
		Also, since $\mbf{x} \in {\kernel(\herm{\mbf{B}})}\pspp$,
		$\herm{\mbf{B}}\pspp \mbf{x}=\mbf{0} 
		\implies \herm{\mbf{B}}\pspp \herm{\mbf{F}}\mbf{y}=\mbf{0} \implies \mbf{y}\in \kernel\big(\herm{(\mbf{F}\pspp\mbf{B})}\big)\pspp$; 
		thus $\exists \ { \mbf{z} \in}\ \mathbb{C}^{q\times 1}_{}$ such that ${\mbf{y}}={\herm{\mbf{G}}\mbf{z}}
		\implies
		{\mbf{x}}={\herm{\mbf{F}}\pspp{\herm{\mbf{G}}\mbf{z}}}\pspp$. 
		Hence, $\mbf{x} \in \range\big(\herm{(\mbf{G}\pspp\mbf{F})}\big)
		\implies  
		{\kernel(\herm{\mbf{A}})}\psp \cap \psp {\kernel(\herm{\mbf{B}})}
		\subseteq\range\big(\herm{(\mbf{G}\pspp\mbf{F})}\big)\pspp$.
		Therefore, $\range\big(\herm{(\mbf{G}\pspp\mbf{F})}\big)
		=
		{\kernel(\herm{\mbf{A}})}\psp \cap \psp {\kernel(\herm{\mbf{B}})}\pspp$; since rows of $\mbf{F}$ and $\mbf{G}$ are linearly independent, the rows of ${\mbf{G}\pspp \mbf{F}}$ are also linearly independent, and hence, the rows of ${\mbf{G}\pspp \mbf{F}}$ form a basis for ${\kernel(\herm{\mbf{A}})}\psp \cap \psp {\kernel(\herm{\mbf{B}})}$.
	\end{proof}
\end{theorem}
%

Motivated by this theorem, we propose the following recursive algorithm for computing of the left nullspace of a block column matrix.

\begin{algorithm}[h]
	\caption{$\!\!\!\nspp$: 
	Recursive estimation of the common left nullspace of a set of matrices
}
	\label{alg:alg1}
	\textbf{Input:} 
	Set of matrices $\mbf{A}_1^{},\,\mbf{A}_{2}^{},\,\ldots,\,\mbf{A}_{L}^{}\pspp$, where $\mbf{A}_i^{}\in\mathbb{C}^{n\times m_i^{}}_{}$, initial left nullspace $\mbf{Z}_{0}^{}$.
	\begin{algorithmic}[1]
		\For {$i\in\mathbb{N}_L^{}$}
		\State $\mbf{T}\gets$  $\mbf{Z}_{i-1}^{}\times\mbf{A}_i^{}$
		\State Construct $\mbf{W}$ whose rows form a basis for $\kernel(\herm{\mbf{T}})$
		\State $\mbf{Z}_{i}^{}\gets {\mbf{W}}  \times \mbf{Z}_{i-1}^{}$
		\EndFor
	\end{algorithmic}
	\textbf{Output:} $\mbf{Z}_{L}^{}\pspp$, whose rows form the basis for 
	$\bigcap\limits_{i \psp \in \psp \mathbb{N}_{L}^{}}^{}\!{\kernel(\hermm{\mbf{A}}_i)}
	\pspp$.
\end{algorithm}

\subsection{Computing Left Nullspace for SD}
We now introduce the proposed approach for decoupling that exploits the redundancy in computing the left nullspaces. 
We can use Algorithm \ref{alg:alg1} to find the common left nullspace of the complementary channel as in \eqref{eq:LNS_intersection}, and we use the intermediate estimates to reduce the computational cost.

Let us consider $K$ 
user channels $\mbf{H}_1^{},\, \mbf{H}_2^{},\, \ldots,\, \mbf{H}_K^{}\pspp$. To begin with, in the first level, we divide the available users into two sets $\mathcal{S}_{1}^{(1)}=\mathbb{N}_{t_{1}^{(1)}}^{}$ and $\mathcal{S}_{2}^{(1)}=\mathbb{N}_{K}^{}\backslash\mathbb{N}_{t_{1}^{(1)}}^{}\pspp$, having indices of  $t_{1}^{(1)}=\big\lceil{\tfrac{{K}}{2}}\big\rceil$ and $t_{2}^{(1)}=
K-t_{1}^{(1)}$ users, respectively. 
The collective channel matrices 
of the users enlisted in $\mathcal{S}_{1}^{(1)}$ and $\mathcal{S}_{2}^{(1)}$ sets are 
\begin{equation*}
	\mbf{H}_{\mathcal{S}_{1}^{(1)}}^{}=\big[\psp\mbf{H}_1^{} \ \cdots \ \mbf{H}_{t_{1}^{(1)}}^{}\pspp\big]
\quad\text{and}\quad
	\mbf{H}_{\mathcal{S}_{2}^{(1)}}^{}=\big[\psp\mbf{H}_{t_{1}^{(1)}+1}^{}  \ \cdots\ \mbf{H}_{K}^{}\pspp\big]
	\psp,
\end{equation*}
respectively. 
Now, we utilize Algorithm \ref{alg:alg1} to compute the left nullspace of $\mbf{H}_{\mathcal{S}_{1}^{(1)}}^{}$ and $\mbf{H}_{\mathcal{S}_{2}^{(1)}}^{}$ as $\mbf{W}_{\mathcal{S}_{1}^{(1)}}^{}$ and $\mbf{W}_{\mathcal{S}_{2}^{(1)}}^{}\pspp$, respectively, which will satisfy the conditions
\begin{equation}
	\mbf{W}_{\mathcal{S}_{1}^{(1)}}^{}\psp \mbf{H}_{\mathcal{S}_{1}^{(1)}}^{}=\mbf{0}
	\quad\text{and}\quad
	\mbf{W}_{\mathcal{S}_{2}^{(1)}}^{} \psp \mbf{H}_{\mathcal{S}_{2}^{(1)}}^{}=\mbf{0}.
\end{equation}
Note that the set $\mathcal{S}_{1}^{(1)}$ contains part of complementary users for each of the user present in the set $\mathcal{S}_{2}^{(1)}$  and vice versa. 
To account for the rest of the complementary users, in the second level, we further divide these sets and find the corresponding left nullspaces;
%
we divide the set in $\mathcal{S}_{2}^{(1)}$ into two sets, $\mathcal{S}_{3}^{(2)}$ and $\mathcal{S}_{4}^{(2)}\pspp$,  
having $t_{3}^{(2)}=\Big\lceil{\tfrac{t_{2}^{(1)}}{2}}\Big\rceil$ and $t_{4}^{(2)}=t_{2}^{(1)}-t_{3}^{(2)}$ users, respectively. 
Similarly, 
we divide the users in $\mathcal{S}_{1}^{(1)}$ into two sets, $\mathcal{S}_{1}^{(2)}$ and $\mathcal{S}_{2}^{(2)}\pspp$, having $t_{1}^{(2)}=\big\lceil{\tfrac{t_{1}^{(1)}}{2}}\big\rceil$ and $t_{2}^{(2)}=t_{1}^{(1)}-t_{1}^{(2)}$ 
users, respectively. 
It is evident that the set $\mathcal{S}_{1}^{(1)}\cup\mathcal{S}_{3}^{(2)}$ comprises a part of complementary users for each of the users in the set $\mathcal{S}_{4}^{(2)}$ and so on. Again, using Algorithm \ref{alg:alg1}, we can compute the left nullspaces of the channel matrices of the users listed in $\mathcal{S}_{1}^{(2)}, \mathcal{S}_{2}^{(2)}$, $\mathcal{S}_{3}^{(2)}$ and $\mathcal{S}_{4}^{(2)}$ $\big($respectively denoted as $\mbf{W}_{\mathcal{S}_{1}^{(2)}}^{}$, $\mbf{W}_{\mathcal{S}_{2}^{(2)}}^{}$, $\mbf{W}_{\mathcal{S}_{3}^{(2)}}^{}$ and $\mbf{W}_{\mathcal{S}_{4}^{(2)}}^{}\big)$, so that  
%
\begin{equation}
	\psp\mbf{W}_{\mathcal{S}_{3}^{(2)}}^{} \psp \mbf{W}_{\mathcal{S}_{1}^{(1)}}^{}\psp\left[\psp\mbf{H}_{\mathcal{S}_{1}^{(1)}}^{}\ \mbf{H}_{\mathcal{S}_{3}^{(2)}}^{}\psp\right]=\mbf{0}
	\psp,	
\end{equation}
\begin{equation}
	\psp\mbf{W}_{\mathcal{S}_{4}^{(2)}}^{}\psp \mbf{W}_{\mathcal{S}_{1}^{(1)}}^{}\left[\psp\mbf{H}_{\mathcal{S}_{1}^{(1)}}^{} \ \mbf{H}_{\mathcal{S}_{4}^{(2)}}^{}\psp\right]=\mbf{0},	
\end{equation}
\begin{equation}\label{eq:S21}
	\psp\mbf{W}_{\mathcal{S}_{1}^{(2)}}^{}\psp \mbf{W}_{\mathcal{S}_{2}^{(1)}}^{}\psp\left[\mbf{H}_{\mathcal{S}_{2}^{(1)}}^{} \ \mbf{H}_{\mathcal{S}_{1}^{(2)}}^{}\right]=\mbf{0},	
\end{equation}
\begin{equation}
	\psp\mbf{W}_{\mathcal{S}_{2}^{(2)}}^{}\psp \mbf{W}_{\mathcal{S}_{2}^{(1)}}^{}\psp\left[\psp\mbf{H}_{\mathcal{S}_{2}^{(1)}}^{}\ \mbf{H}_{\mathcal{S}_{2}^{(2)}}^{}\psp\right]=\mbf{0}.
\end{equation}
Thus, the intermediate left nullspaces $\mbf{W}_{\mathcal{S}_{1}^{(1)}}^{}$ and $\mbf{W}_{\mathcal{S}_{2}^{(1)}}^{}$ from the previous level can be used to initialize Algorithm \ref{alg:alg1} to compute the left nullspaces at the second level. 

Proceeding in this way, we continue until there are subsets that cannot be divided further, i.e., until $\mathcal{S}_{1}^{(1)}$ gets partitioned into $\big\lceil{\tfrac{{K}}{2}}\big\rceil$ subsets and  $\mathcal{S}_{2}^{(1)}$ 
into $K-\big\lceil{\tfrac{{K}}{2}}\big\rceil$ subsets.
This will result in $\big\lceil{\log_2{K}}\big\rceil$ 
	levels, and finally, there will be $K$ left nullspaces, which constitute the decouplers of each of the $K$ users. Indeed, the decoupler thus derived for each user will satisfy 
$
	\mbf{W}_{i}^{}
	\left[\pspp \mbf{H}_K^{}\ \cdots\ \mbf{H}_{i+1}^{} \ \mbf{H}_{i-1}^{} \cdots \ \mbf{H}_{1}^{}\right]
	=\mbf{0}\pspp,
$ 
where $\mbf{W}_{i}^{}\in \mathbb{C}^{\overline{M}_i^{}\times N_R}$ helps parallel processing of every user.
Hence, this approach provides a low-complex linear decoupler 
by using intermediate left nullspace estimates at each level; since the left nullspaces are computed sequentially,  the decoupler is named so. We summarize the above logic to obtain the decouplers in Algorithm \ref{alg:alg2}.
\begin{algorithm}
	\caption{$\!\!\!\nspp\ $: Sequential Decoupler (SD)}
	\label{alg:alg2}
	\vspace{0.25ex}
	\textbf{Input:} 
	User channels $\mbf{H}_i^{}\in \mathbb{C}^{N_R^{}\times M_i^{}}_{},\ i\in \mathbb{N}_K^{}\pspp$.
	\\[0.75ex]
	\textbf{Initialization:} 
			Number of levels,
			$\nu=\lceil\psppp{\log_2^{}{K}}\psppp\rceil\pspp$, left nullspace initialization $\mbf{Z}_{1}^{(0)}=\mbf{I}_{N_R^{}}\pspp$, 
					$\mathcal{A}_{1}^{(0)}=\emptyset, \
					\mathcal{B}_{1}^{(0)}=\mathbb{N}_K^{}\pspp
					$. 
	\begin{algorithmic}[1] 
		\For{$l\in 
				\mathbb{N}^{}_{\pspp\nu}$}
		\vspace{0.5ex}
		\For{$i\in \mathbb{N}_{\pspp2^{l-1}}$}
		\vspace{0.6ex}
		\State $\xi\gets\big|\mathcal{B}_{i}^{(l)}\big| 
			,
			\,\
			\mathcal{C}_1^{}\gets\mathbb{N}^{}_{\lceil\psppp{{\xi}/{2}}\psppp\rceil},
			\,\
			\mathcal{C}_2^{}\gets\mathbb{N}_{\xi}^{}\big\backslash\,\mathcal{C}_1^{}$
		\vspace{0.6ex}
		\For{$j\in \mathbb{N}_{\pspp2}^{}$}
		\vspace{0.5ex}
		\State $\lambda\gets2\pspp(i-1)+j$
		\State 
		\hspace{-0.65em}\begin{tabular}{L{19.5em}}
			Get $\mbf{Z}_{\pspp\lambda}^{(l)}\,$ using Algorithm \ref{alg:alg1} with 
				input $\mbf{H}_p^{},$\\[-0.8ex]
			$p\in \mathcal{B}_{i}^{(l-1)}\nspp[\psp\mathcal{C}_{3-j}^{}\psp]\pspp$,
				and initialization $\mbf{Z}_{i}^{(l-1)}\pspp$.
		\end{tabular}
		\vspace{0.5ex}
		\State 
		$\mathcal{A}_{\pspp\lambda}^{(l)}=\mathcal{A}_{i}^{(l-1)}\cup\psppp\mathcal{C}_{3-j}^{}$
		\vspace{1ex}
		\State $\mathcal{B}_{\pspp\lambda}^{(l)}=\mathcal{B}_{i}^{(l-1)}\nspp[\psp\mathcal{C}_{j}^{}\psp]$
		\vspace{0.35ex}
		\EndFor
		\vspace{0.25ex}
		\EndFor
		\vspace{0.25ex}
		\EndFor 
		\State 
		\hspace{-0.5em}\begin{tabular}{L{22em}}
			For every $i\in\{j\in\mathbb{N}_{\pspp 2^{\nu}_{}}^{} \,|\, \mathcal{B}_i^{(\nu)}\nsp\neq\emptyset\}\pspp$, assign $\mbf{W}_{p_i^{}}^{}$ with $\mbf{Z}_i^{(\nu)}$, where $p_i^{}=\mathcal{B}_i^{(\nu)}[\{1\}]$
			\end{tabular}
			\vspace{0.5ex}
	\end{algorithmic}
	\vspace{0.25ex}
	\textbf{Output:} 
		Decouplers of all users, $\mbf{W}_i^{},\ i\in\mathbb{N}_{K}^{}\pspp$.
\end{algorithm}
%
%
%
%
\subsection{Users Inclusion}{\label{UsersAddition}
We now discuss how the proposed SD is adaptable to the inclusion of new users into the system. 
When new users are added, the existing SD can be updated with less number of computations, 
unlike other
schemes 
\cite{arevalo2015uplink,arevalo2017decoupled,zu2019uplink} 
that require the re-computation of the entire decoupler. 
Let us consider there are already $K$ users 
in the system and let ${P}$ denote the number of users wanting to be newly added to the system. 
%
%
Algorithm \ref{alg:alg3} illustrates how the new set of decouplers can be determined from the existing ones with less computational cost rather than computing decouplers for the entire set of users. 
The Algorithm derives the decoupler of a new user 
from that of an existing user. 
The decouplers of all the existing users are then updated by accounting for the new user's channel, 
}
\begin{algorithm}[h]
	\caption{$\!\!\!\nspp\ $: Users inclusion in 
			Sequential Decoupler, SD\hyph UI}
	\label{alg:alg3}
	\vspace{0.5ex}
	\textbf{Input:} Channels of existing users 
	$\mbf{H}_i^{},\ i\in \mathbb{N}_K^{}$, 
	decouplers  of existing users 
	$\mbf{W}_i^{},\ i\in \mathbb{N}_K^{}$ 
	and channels of new users 
	$\mbf{H}_{i}^{},\ i\in\mathbb{N}_{{P}}^{}+K\pspp$.
	\\[0.75ex]
	\textbf{Initialization:} Indices of users with decouplers, $\mathcal{U}=\mathbb{N}_K^{}\pspp$.
	\begin{algorithmic}[1]
		\For {$i\in\mathbb{N}_{{P}}^{}$}
		\vspace{0.5ex}
		\State Pick an arbitrary index 
		$p\in\mathcal{U}$
		\vspace{-0.75ex}
		\State 
		\hspace{-0.65em}\begin{tabular}{L{22em}}
			Get $\mbf{W}_{K+i}^{}$ using Algorithm \ref{alg:alg1} with input $\mbf{H}_p^{}$ and initialization $\mbf{W}_p^{}\pspp$.\\[3.75ex]
		\end{tabular}
		\For {$j\in\mathcal{U}$}
		\vspace{-0.5ex}
		\State 
			\hspace{-0.65em}\begin{tabular}{L{21em}}
			Replace $\mbf{W}_{j}^{}$ with output of Algorithm \ref{alg:alg1} with input $\mbf{H}_i^{}$ and initialization $\mbf{W}_j^{}\pspp$.\\[4ex]
		\end{tabular}
		\EndFor
		\vspace{0.25ex}
		\State $\mathcal{U} \gets \mathcal{U} \cup \{i\}$
		\vspace{0.25ex}
		\EndFor
	\end{algorithmic}
	\vspace{0.25ex}
	\textbf{Output:} 
	Decouplers of all users, $\mbf{W}_i^{},\ i\in\mathbb{N}_{K+{P}}^{}\pspp$.
\end{algorithm}
%
%
%
%
\begin{table*}[ht]
	\caption{Computational complexity (in terms of FLOPs) of the Decoupler\hyph SVD \cite{arevalo2015uplink} and the proposed Sequential Decoupler}
	\label{tab:Complexity_Table}
	\begin{subtable}{\linewidth}
		\caption{with varying numbers of users, each user having ${M}_i^{}=2$ antennas}
		\label{tab:Complexity_users}
		\centering
		\begin{tabular}{?c?C{5em}|C{5em}|C{5em}|C{5em}|C{5em}|C{5em}?}
			
			\specialrule{1pt}{0pt}{0pt}
			\multirow{2}{2cm}{\centering\textbf{Scheme}} & \multicolumn{6}{c?}{\textbf{No. of users,} $K$}\\
			\cline{2-7}
			& 30 & 40 & 50 & 60 & 70 & 80\\
			\specialrule{1pt}{0pt}{0pt}
			\specialrule{0.8pt}{1pt}{0pt}
			
			Decoupler\hyph SVD \cite{arevalo2015uplink} & 568381440 & 1837854720 & 4548454400 & 9517332480 & 17745960960 & 30420131840\\
			\hline 
			 Proposed SD & 2350036 & 5677832 & 11180738 & 19385412 & 30947856 & 46378784\\
			
			\specialrule{1pt}{0pt}{0pt}
		\end{tabular}	
	\end{subtable}
	\vskip 0.1in
	\begin{subtable}{\linewidth}
		\caption{with varying number of streams per user when $K=50$}
		\label{tab:Complexity_ant}
		\centering
		\begin{tabular}{?c?C{6em}|C{6em}|C{6em}|C{6em}?}
			
			\specialrule{1pt}{0pt}{0pt}
			\multirow{2}{2cm}{\centering\textbf{Scheme}} & \multicolumn{4}{c?}{${M}_i^{}$}\\
			\cline{2-5}
			& 2 & 4 & 6 & 8\\
			
			\specialrule{1pt}{0pt}{0pt}
			\specialrule{0.8pt}{1pt}{0pt}
			Decoupler\hyph SVD \cite{arevalo2015uplink} & 4548454400 & 3638763520 & 122808268800 & 291101081600\\
			\hline 
		 Proposed SD & 11180738 & 89283324 & 301141166 & 713587672 \\
			
			\specialrule{1pt}{0pt}{0pt}
		\end{tabular}
	\end{subtable}
\end{table*}
\begin{figure*}[ht]
	\centering
	\begin{subfigure}[b]{0.48\linewidth}
		\centering
		\includegraphics[width=1\linewidth]{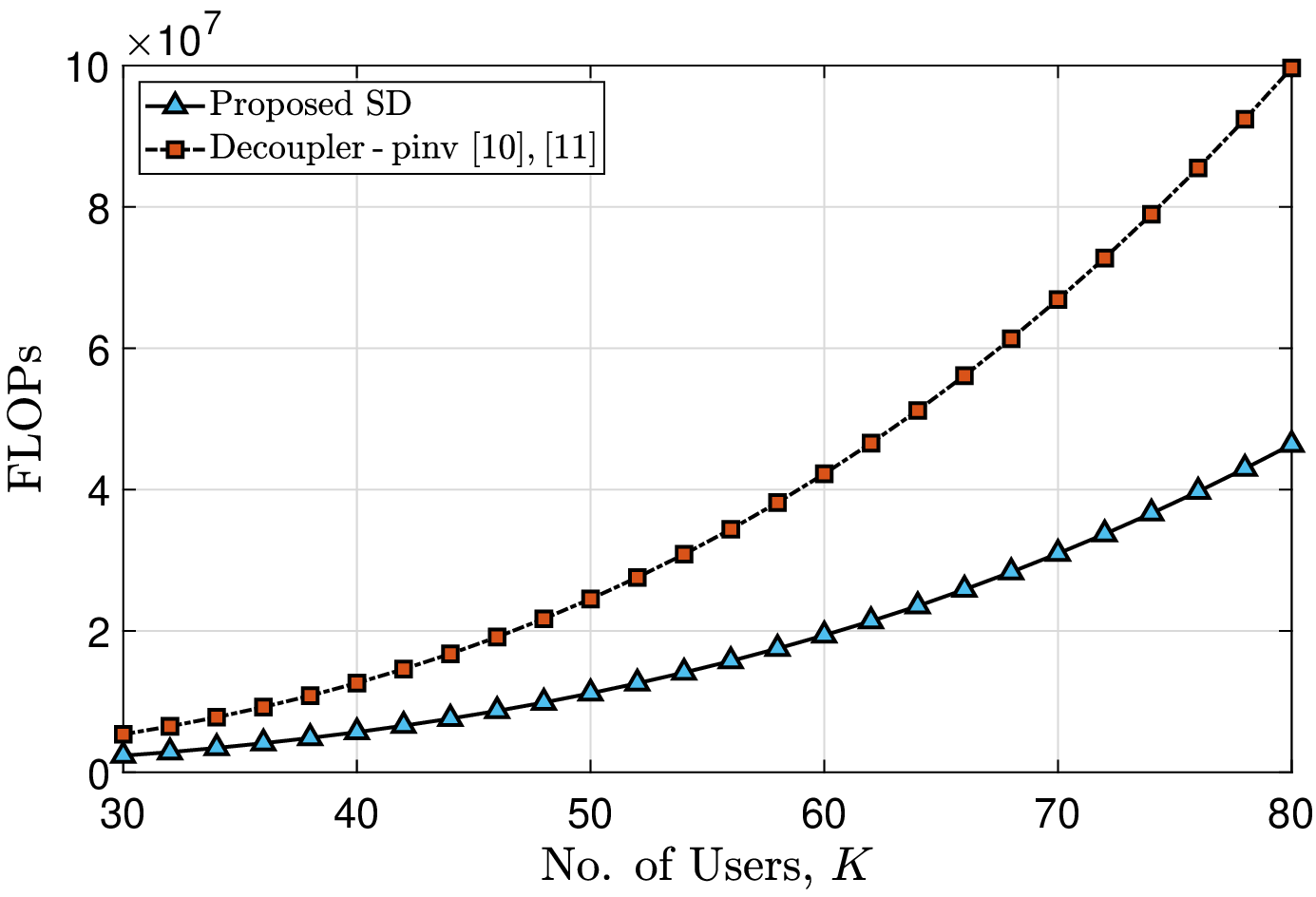}
		\caption{with varying numbers of users, each user having ${M}_i^{}=2$ antennas}
		\label{fig:complexity_users}
	\end{subfigure}
	\hspace{0.027\linewidth}
	\begin{subfigure}[b]{0.48\linewidth}
		\centering
		\includegraphics[width=1\linewidth]{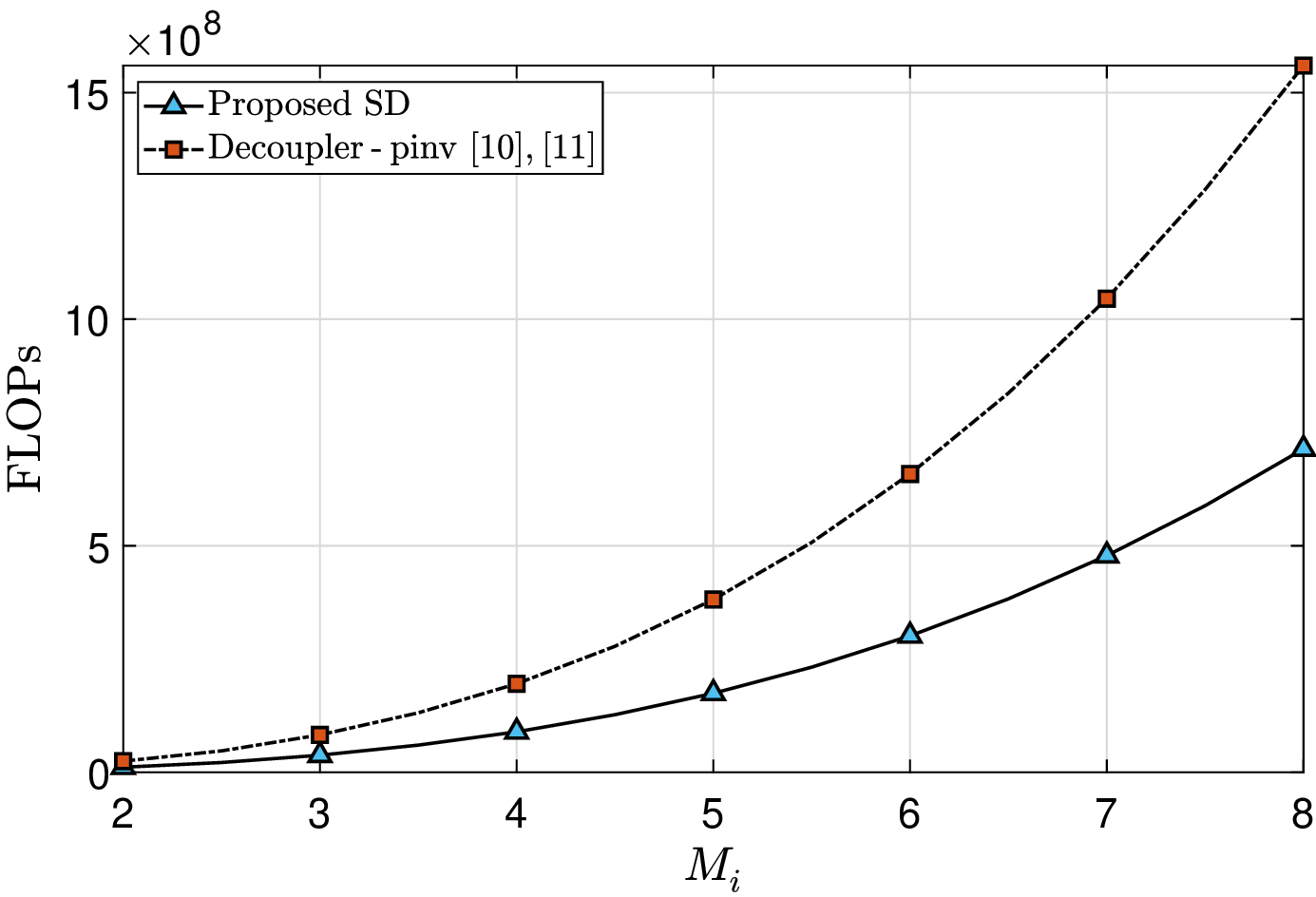}
		\caption{with varying number of streams per user when $K=50$}
		\label{fig:complexity_antennas}
	\end{subfigure}
	\caption{Computational complexity (in terms of FLOPs) of the Decoupler\hyph pinv \cite{arevalo2017decoupled,zu2019uplink} and the proposed SD.}
	\label{fig:Compt_Complexity}
\end{figure*}
\begin{table*}
	\caption{Computational Complexity (in terms of FLOPs) of the schemes when up to five users (with $M_i^{}=2$) get added to $\{130, 60 , 2\}$ system.}
	\label{tab:tab3}
	\centering
	\begin{tabular}{?c?C{5em}|C{5em}|C{5em}|C{5em}|C{5em}?}
		
		\specialrule{1pt}{0pt}{0pt}
		\multirow{2}{2cm}{\centering\textbf{Scheme}} & \multicolumn{5}{c?}{Additional number of users, $P$}\\
		\cline{2-6}
		& 1 & 2 & 3 & 4 & 5\\
		
		\specialrule{1pt}{0pt}{0pt}
		\specialrule{0.8pt}{1pt}{0pt}
		Decoupler\hyph SVD \cite{arevalo2015uplink} & 10400256000 & 11044166144 & 11717250048 & 12420366336 &13154385920\\
		\hline 
		Decoupler\hyph pinv \cite{arevalo2017decoupled,zu2019uplink} & 46323278 & 48086828 & 49890834 & 51735680 & 53621750\\
		\hline
		Sequential Decoupler (SD) & 28914457 & 28034542 & 27030719 & 25899264 & 24673804 \\
		\hline
		Sequential Decoupler$\hyph$Users Inclusion & 26975155 & 4917261 & 6702870 & 8008150 & 8808885 \\
		\specialrule{1pt}{0pt}{0pt}
	\end{tabular}
\end{table*}
\section{Empirical Validations}
In this section, we compare the performance of the proposed SD with existing decoupling schemes 
 Decoupler\hyph pinv \cite{arevalo2017decoupled}, \cite{zu2019uplink}
   and Decoupler\hyph SVD \cite{arevalo2015uplink} through numerical simulations in terms of computational complexity and Bit Error Rate (BER). Throughout the section, we use the notation $\{N_R^{},K,M_i^{}\}$ to denote the number of receiver antennas at the BS, the number of users, and the number of transmit antennas per user, respectively.
\subsection{Computation Complexity}
We compare the computational complexity of the algorithms in terms of Floating Point Operations (FLOPs)\cite{golub2013matrix}.
In Table \ref{tab:Complexity_Table}, the complexity comparison between the proposed SD approach and  Decoupler\hyph SVD is presented; Table \ref{tab:Complexity_users} furnishes the computational complexity when the number of users increases; Table \ref{tab:Complexity_ant} 
provides the computational complexity when the number of transmit antennas increases for all users. 
It is  evident that the proposed SD has substantially reduced the computational complexity.  
A key result is that when the number of users in the system increases from 30 to 80, the computational complexity required by the proposed SD approach decreases from $0.41\%$ to $0.152\%$. 
Figures \ref{fig:complexity_users} and \ref{fig:complexity_antennas} plot the FLOPs required by our SD approach and Decoupler\hyph pinv.
We can clearly see that the proposed 
decoupler has remarkably reduced the computational complexity and only requires $46\%$ of the FLOPs as Decoupler\hyph pinv. 
%

From the above results, it is apparent that the proposed scheme offers a significant 
reduction in the number of computations over the state-of-the-art, helping to achieve fast detection. 
%
%
In particular, the complexity reduction is more pronounced when the dimension of the system increases. This makes the scheme apt for handling a large number of users,  which is typically the case in massive MIMO systems. 

%
%
%
%
%
%
%
%

	Next, we study the computational complexity for adapting the Sequential Decoupler when new users are included, i.e., SD\hyph UI discussed in 
Algorithm \ref{alg:alg3}. 
Table \ref{tab:tab3} shows the 
	FLOPs required 
	when up to five users are included, starting with $K=60$. 
	Note that Decoupler\hyph SVD 
	and
	Decoupler\hyph pinv 
	require the decoupler to be computed for all users again 
	and, hence, are not scalable.
	For reference, we have also provided the FLOPs required to compute the entire SD again from scratch.
	The results elucidate that the computational complexity can be drastically reduced  by cleverly harnessing the hierarchical structure of 
	SD. 
	Notably, when five users are added to the system, the SD\hyph UI scheme requires only $0.0670\%$ and $16.42\%$ of the FLOPs needed for Decoupler-SVD and  Decoupler-pinv, respectively; this amounts to $35.7\%$ of the FLOPs required for computing the SD from scratch. 
	%
\subsection{Bit Error Rate (BER) Analysis}
This section provides BER analysis of the decouplers; after decoupling, user streams are detected using schemes mentioned in Section \ref{Detection_schems}. We first study the performance under channel effects such as small-scale fading, spatial correlation, large-scale propagation effects, and channel impairments such as channel estimation error (CE).
	\subsubsection{Kronecker Model \cite{chockalingam2014large}}
	We define a correlated channel model according to Kronecker model \cite{chockalingam2014large}. \begin{equation}
		\mbf{H}_{i}^{}=\left(\mbf{S}_{\text{rx}}^{}\right)^{\frac{1}{2}} \psp \widecheck{\mbf{H}}_{i}^{} \psp \left(\mbf{S}_{\text{tx}}^{}\right)^{\frac{1}{2}}
	\end{equation}
	where the entries of $\widecheck{\mbf{H}}_{i}^{}$ are i.i.d. samples from $\mathcal{CN}(0,1)$; $\mbf{S}_\text{rx}^{}$ is the correlation matrix at the BS and $\mbf{S}_\text{tx}^{}$ is the correlation matrix at the user terminals, 
	both of which are constructed as
	\begin{equation}
		\mbf{S}_{a}^{}=\begin{bmatrix}
			1 & \rho_a^{} &  \ldots & \rho_a ^{(N_a^{}-1)^{2}}\\
			\rho_a^{} & 1 & \ldots & \rho_a^{(N_a^{}-2)^{2}}\\
			\vdots & \vdots &\ddots &\vdots\\
			\rho_a ^{(N_a-1)^{2}}& \rho_a^{(N_a^{}-2)^{2}} & \ldots  & 1\\
		\end{bmatrix}
		\pspp, 
	\end{equation}
	where $a \equiv \text{`tx' or `rx'}\pspp$, $N_{\text{tx}}^{}=M_i^{}$ and $N_{\text{rx}}^{}=N_R^{}\pspp$. We have used $\rho_{\text{tx}}^{}=0.25$ and $\rho_{\text{rx}}^{}=0.05$ for the simulations.
	Figure \ref{fig:BER_uncorrelated} shows the BER performance in a $\left\{{64,15,4}\right\}$ system 
	under 
	the uncorrelated channel (i.e., $\rho_{\text{tx}}^{}=\rho_{\text{rx}}^{}=0$), and Figure \ref{fig:BER_correlated} shows the performance of the channel in the presence of correlation. It is clearly evident that the proposed SD does not undergo any performance degradation despite the sequential nature of the proposed algorithm.
	
	\par 
	Now, we consider large-scale propagation effects such as shadowing and 
	path loss in the channel as 
	$\mbf{\Gamma}_i^{}\psp\widecheck{\mbf{H}}_{i}^{}\pspp$, where
	$
		\mbf{\Gamma}_i=\diag\left(10^\frac{\mu_i^{}\nu_i^{}}{10}\sqrt{\tfrac{L_i^{}}{d_i^\tau}}\right),
	$ 
	$\mu_i^{}$ is the shadowing spread in decibels, $\nu_i^{}\sim \mathcal{N}(0,1)$, $L_i^{}$ is the power path-loss between user and the BS and $d_i^\tau$ is the relative distance between the user and the BS with path-loss component $\tau$.
	The parameters of the  simulation are $\mu_i=3\pspp$dB, $L_i^{}=0.65$, $d=0.65\pspp$, and $\tau=3$. For simplicity, all the users are assumed to have the same large-scale parameters.
	The results are provided in Figure \ref{fig:BER_correlated_large_scale_prop}; results show that the presence of large-scale propagation effects cannot deteriorate the performance of the proposed SD.
\subsubsection{Performance Evaluation in the presence of Channel Estimation Error}
When the BS does not have perfect channel knowledge 
(for instance, when there is a channel estimation error), the channel available to the BS can be modelled as \cite{stankovic2004multi},
\begin{equation}{\label{eq:CE}}
	\mbf{H}_{\text{err},i}^{}=\mbf{H}_{i}^{}+\Delta\mbf{H}_i^{}
	\pspp,
\end{equation}
where $[\Delta\mbf{H}_i^{}]_{m,n}^{} 
\!\nsp
\begin{array}{c}
	\\[-2.95em]
	\text{\scriptsize{i.i.d.}}
	\\[-1.25em]
	\sim
\end{array}
\mathcal{CN}\pspp(\mbf{0},  \sigma_{e}^2)\pspp$, $m\in\mathbb{N}_{N_R^{}}\pspp$, $n \in \mathbb{N}_{M_i^{}}^{}\pspp$.
%
The BER performances of various schemes in the presence of a correlated channel under such erroneous channel knowledge are plotted in Figure  \ref{fig:BER_CE}.
It can be observed that our SD scheme performs on par with others as before; therefore, 
the sequential process of the proposed algorithm does not propagate the channel estimation error of a user to affect the quality of decoupling of the other users. 
\begin{figure*}[ht]
	\centering
	\begin{minipage}{0.48\linewidth}
		\begin{center}
			\centerline{\includegraphics[width=\columnwidth]{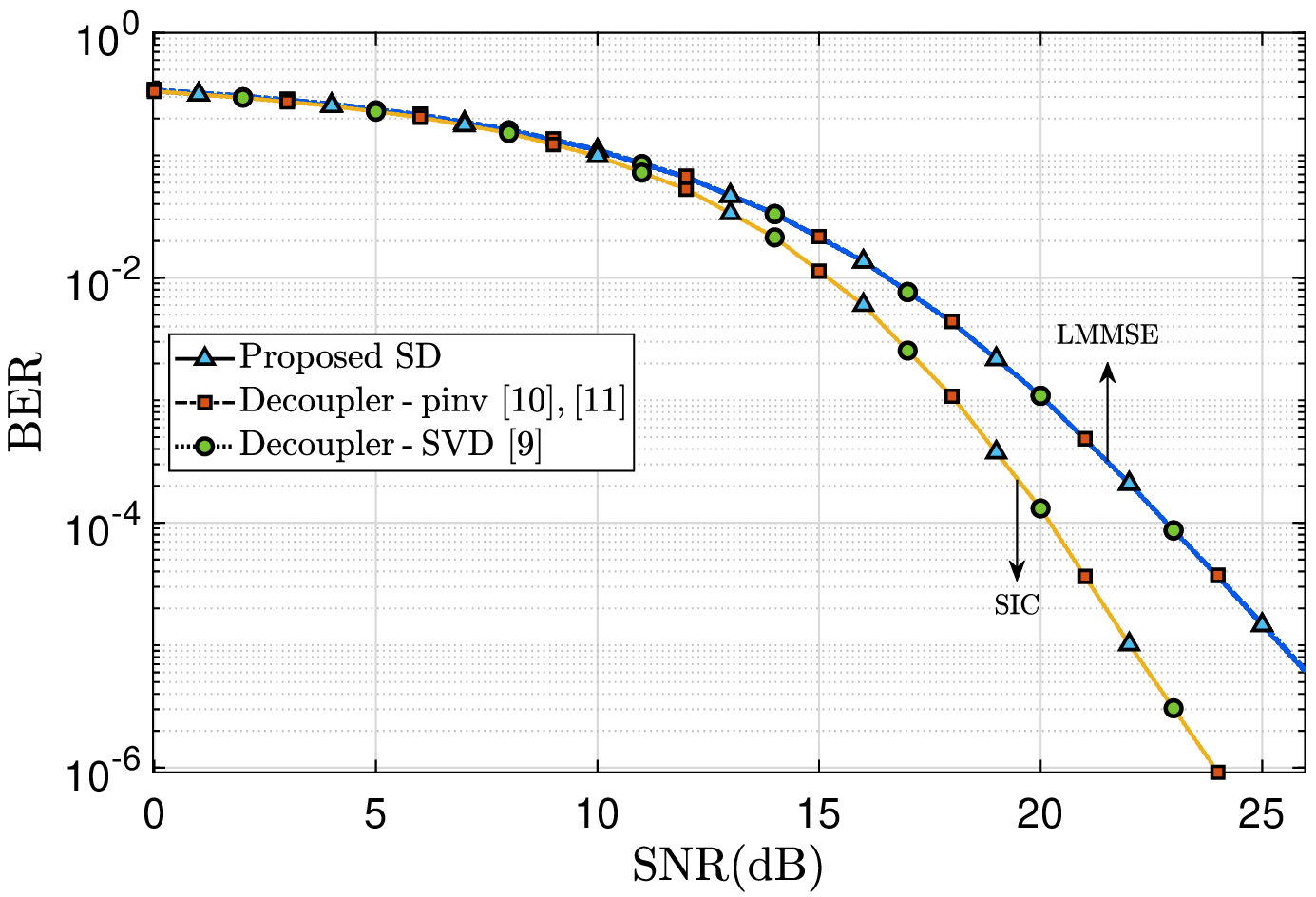}}
			\caption{BER performance in a $\{64,15,4\}$ system under uncorrelated channel}
			\label{fig:BER_uncorrelated}
		\end{center}
	\end{minipage}%
	\hspace{0.027\linewidth}
	\begin{minipage}{0.48\linewidth}
		\begin{center}
			\centerline{\includegraphics[width=\columnwidth]{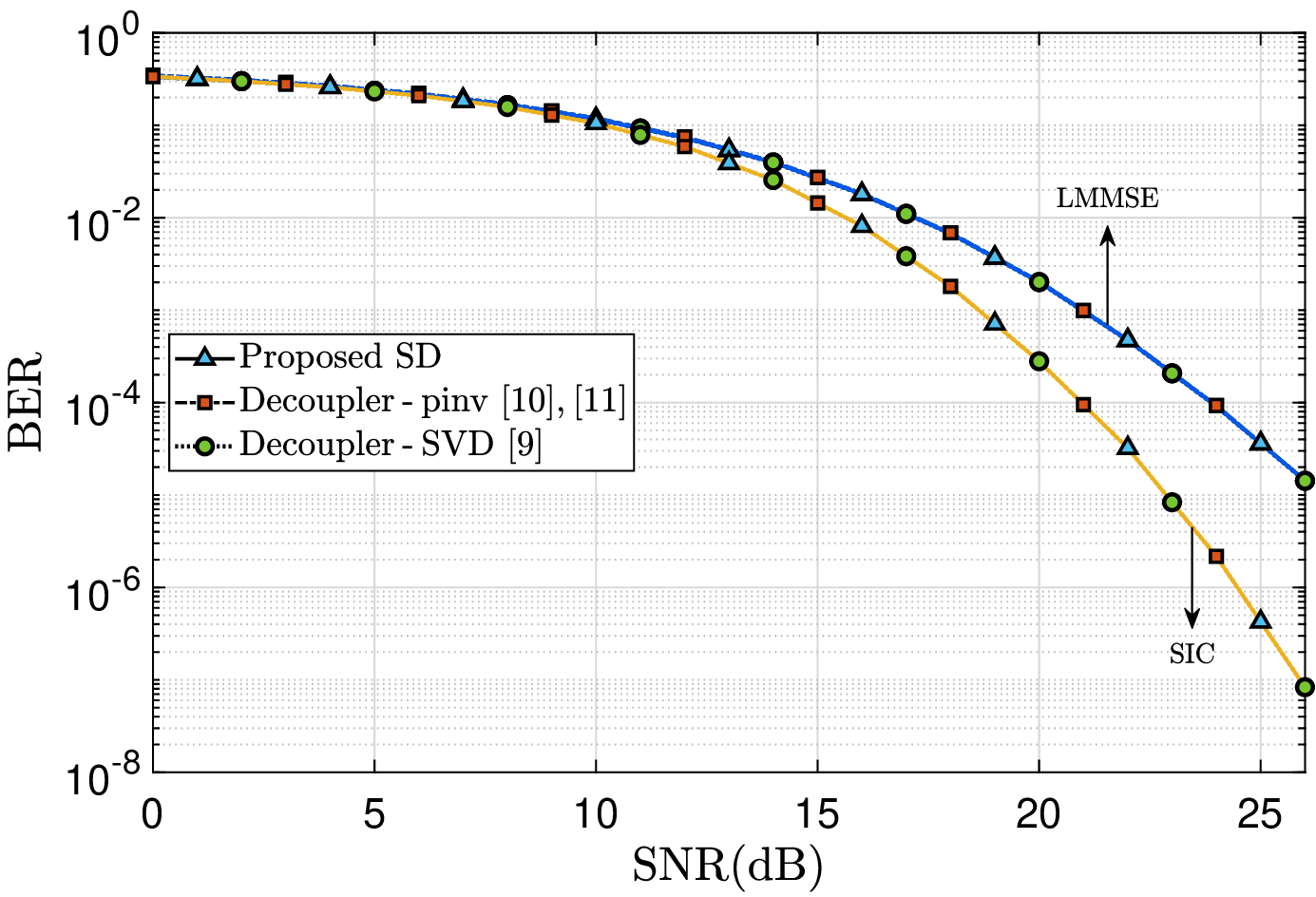}}
			\caption{BER performance in a $\{64,15,4\}$ system under correlated channel with $\rho_{\text{tx}}^{}=0.25$ and $\rho_{\text{rx}}^{}=0.05$}
			\label{fig:BER_correlated}
		\end{center}
	\end{minipage}
\end{figure*}
\begin{figure*}[ht]
	\centering
	\begin{minipage}{0.48\linewidth}
		\begin{center}
			\centerline{\includegraphics[width=\columnwidth]{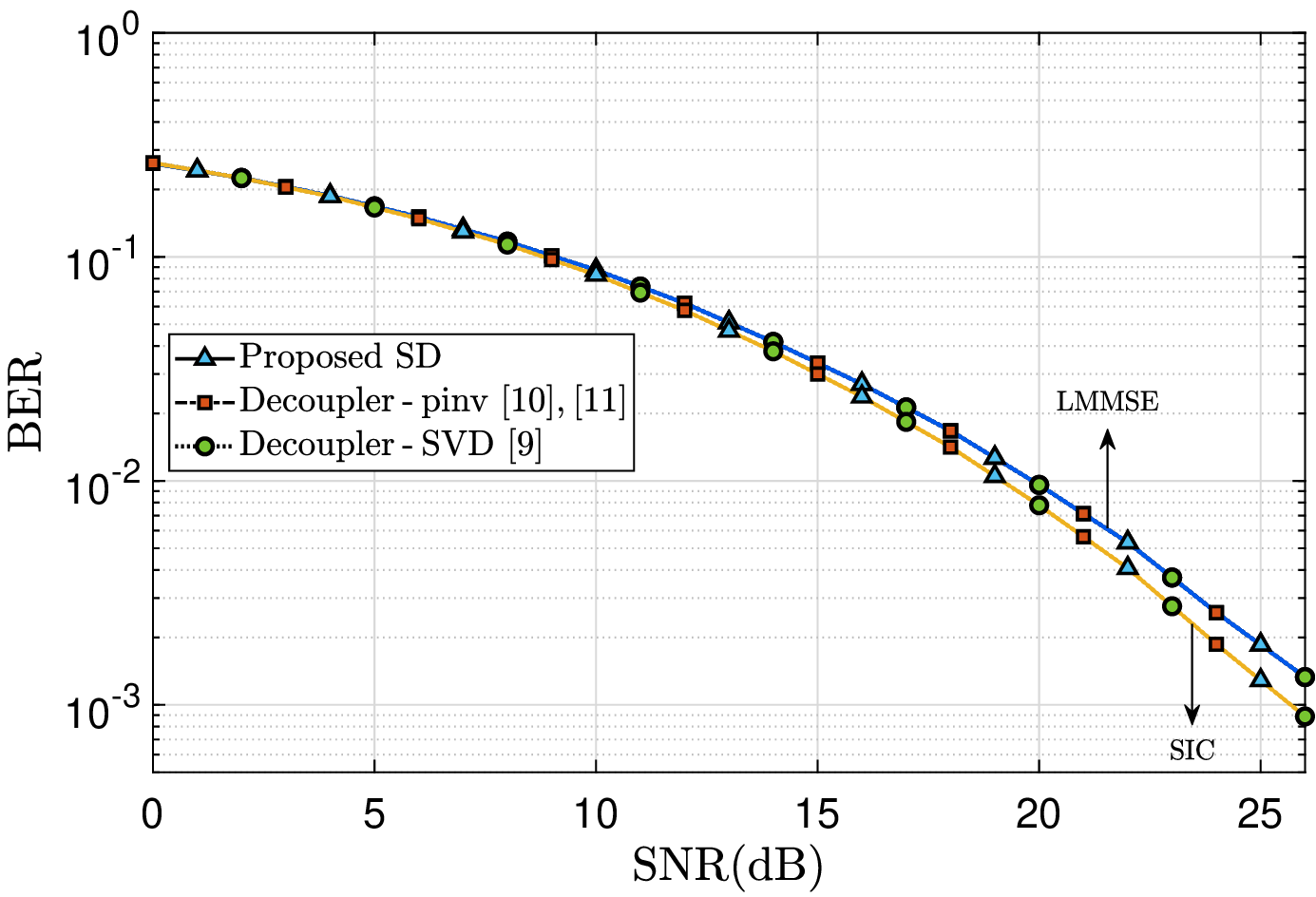}}
			\caption{BER performance in a $\{32,15,2\}$ system under 
				large-scale propagation effects}
			\label{fig:BER_correlated_large_scale_prop}
		\end{center}
	\end{minipage}%
	\hspace{0.027\linewidth}
	\begin{minipage}{0.48\linewidth}
		\begin{center}
			\centerline{\includegraphics[width=\columnwidth]{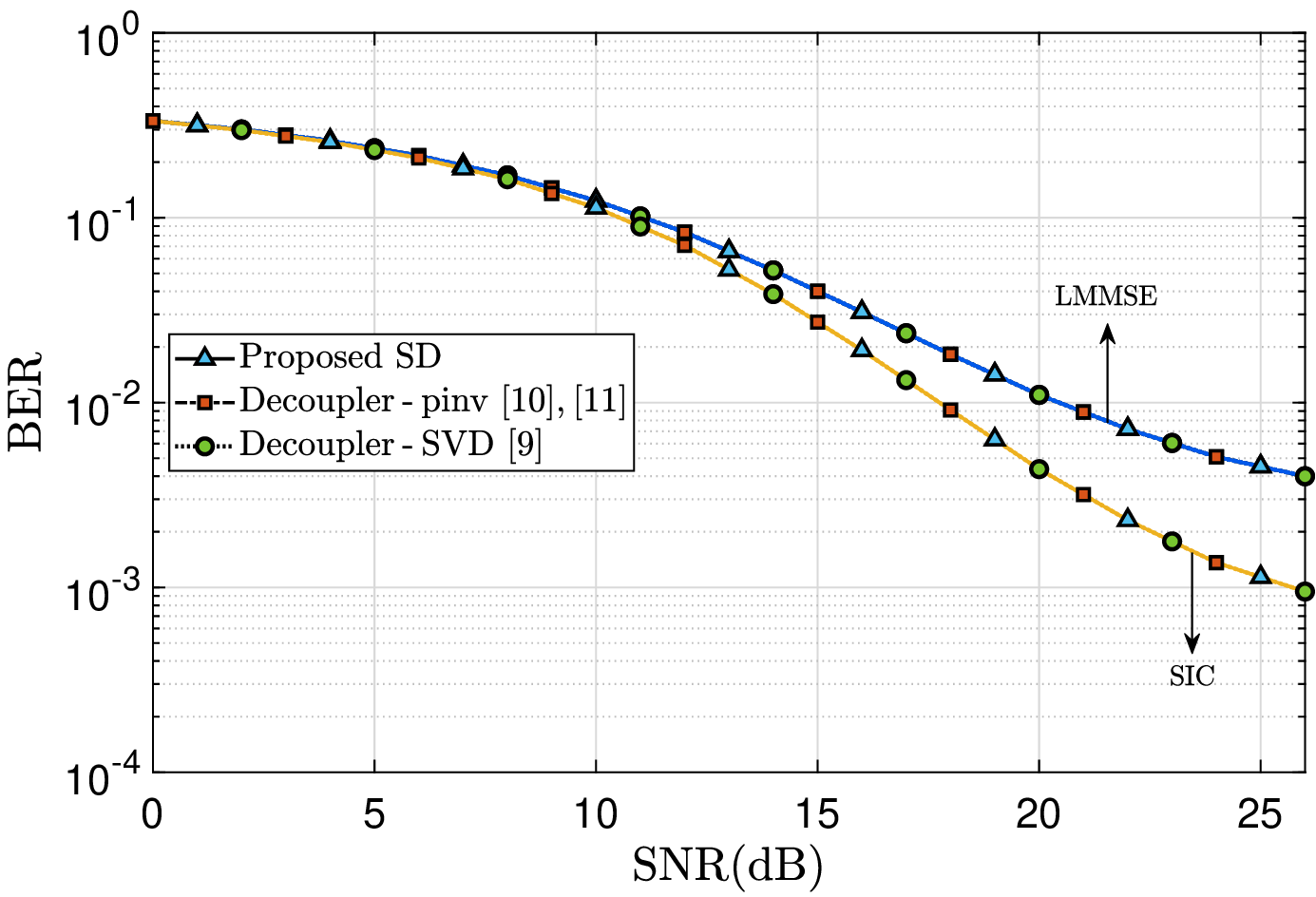}}
			\caption{BER performance in a $\{64,15,4\}$ system in the presence of channel estimation error}
			\label{fig:BER_CE}
		\end{center}
	\end{minipage}
\end{figure*}
\begin{figure}
\centering{\includegraphics[width=\columnwidth]{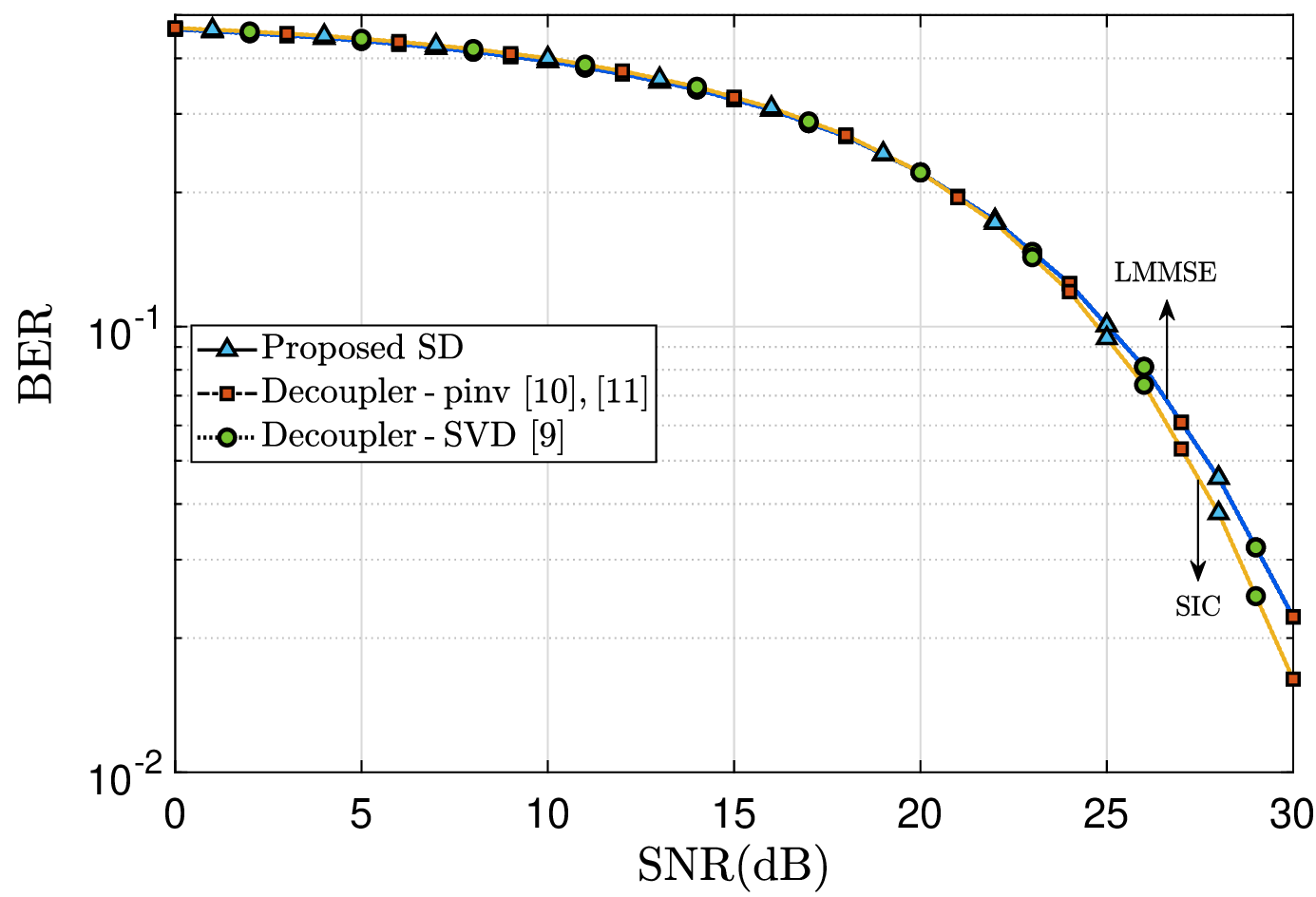}}
\caption{BER performance in a $\{32,15,2\}$ system under TDL-A channel}
\label{fig:BER_TDL_A}
\end{figure}

\subsubsection{TDL-A channel model of 38.901 \cite{TS_38_901}}
We now study the BER performance of the decouplers in an MU-mMIMO-OFDM system, where the frequency selective MIMO channels become flat over each sub-carrier. Thus, decoupling needs to be performed on each sub-carrier, and hence, the proposed decoupler will be desirable in such large-scale systems due to its attractive complexity reduction.
We consider the Tapped Delay Line (TDL-A) channel model, defined in 3GPP Technical Specification 38.901 \cite{TS_38_901}, with parameters listed in  Table \ref{tab:tab4}. 
%
%
%
Figure \ref{fig:BER_TDL_A} shows the BER analysis of a $\left\{{32,15,2}\right\}$ system.
It is evident that the proposed decoupler does not undergo any performance degradation 
and performs on par with the existing schemes.
It corroborates the fact that our SD is a superior alternative to the existing decoupling schemes. 
\begin{table}{}
	\renewcommand{\arraystretch}{0.6}
	\caption{Channel parameters for simulations on the TDL-A model.}
	\label{tab:tab4}
	\centering
	\begin{tabular}{?C{13em}|C{10em}?}
		\specialrule{1pt}{0pt}{0pt}
		\hline
		\textbf{Parameters} & \textbf{Value}\\
		\specialrule{1pt}{0pt}{0pt}
		\specialrule{0.8pt}{1pt}{0pt}
		Delay Profile & TDL-A\\
		\hline
		Subcarrier Spacing & $15\pspp$kHz\\
		\hline
		Resource Blocks & 25\\
		\hline
		Sample Rate &  $7.86\times 10^6$\\
		\hline
		Number of Subcarriers & 300\\ 
		\hline 
		Number of Users & 15\\
		\hline
		$M_i^{}$ & 2\\
		\hline
		$N_R^{}$ & 32\\
		\hline
		SNR ({dB}) & $0\ \text{to}\ 26$ \\
		\hline
		Modulation & QPSK\\
		\hline
		Type of Users Accomodated & Humans, Cars, Bicycle\\
		\hline
		Maximum	Doppler Shift & $163\pspp${Hz}\\
		\specialrule{1pt}{0pt}{0pt}
	\end{tabular}
\end{table}
\section{Conclusions and Future Directions}
%
%
In this paper, we described a novel, computationally efficient linear decoupler for the MU-mMIMO system serving heterogeneous users in the uplink. 
The proposed sequential decoupler 
avoids redundant computations and offers significant complexity reduction by harnessing the appropriate intermediate estimates. 
By cleverly leveraging the hierarchical structure, the proposed SD can be efficiently scaled when new users get added to the system. 
Extensive numerical simulations indicate that the proposed scheme obtains the decoupler with a remarkably low computational cost without any degradation in the performance. 
Future research could consider a regularized version of the Sequential Decoupler to effectively manage the noise in the system.
\bibliographystyle{IEEEtran}
\bibliography{references.bib}
\end{document}